\documentclass[aps,prb,twocolumn]{revtex4-1}
\usepackage{graphicx,color}% Include figure files
\usepackage{dcolumn}% Align table columns on decimal point
\usepackage{bm}% bold math
\usepackage{amsmath,amsthm,amssymb}
\usepackage{gt13}

\begin{document}

%Title of paper 
\title{
Thermal vector potential theory of magnon-driven magnetization dynamics 
}

\author{Gen Tatara}
%\email[johe@physnet.uni-hamburg.de]{}
%\homepage[]{Your web page}
%\thanks{}
\affiliation{%
 RIKEN Center for Emergent Matter Science (CEMS)\\  
2-1 Hirosawa, Wako, Saitama 351-0198, Japan
}

\begin{abstract}
Thermal vector potential formulation is applied to study thermal dynamics of magnetic structures in insulating ferromagnets. 
By separating variables of the magnetic structure and magnons, the equation of motion for the structure including spin-transfer effect due to thermal magnons is derived in the case of a domain wall and a vortex. 
The magnon current is evaluated based on a linear response theory with respect to the thermal vector potential representing the temperature gradient.
It is shown that the velocity of a domain wall when driven by thermal magnon has a strong temperature dependence unlike the case of an electrically-driven domain wall in metals. 
\end{abstract}

\date{\today}

%%%%%%%%%%%%%%%%%%%%%%%%%%%%%%%%%%%%%
\maketitle

\newcommand{\alphaG}{\alpha_{\rm s}}
\renewcommand{\betana}{\beta_{\rm s}}

\newcommand{\Jl}{J_0}
\newcommand{\Dl}{D_0}
\newcommand{\al}{a}
\newcommand{\boson}{b}
%%%%
\newcommand{\Aad}{A_{\rm s}}
\newcommand{\Aadv}{\Av_{\rm s}}
\newcommand{\AU}{A_U}
\newcommand{\AUv}{\Av_U}
\newcommand{\epk}{\epsilon_{k}}
\newcommand{\gammak}{\gamma_k}
\renewcommand{\gyro}{\gamma_{\rm g}}
\newcommand{\HAT}{H_{A_T}}
\newcommand{\Hdwsw}{H^{\rm st}_J}
\newcommand{\HstAT}{H^{\rm st}_{A_T}}
\renewcommand{\Hst}{H_{\rm st}}
\newcommand{\intrv}{\int{d^3r}}
\newcommand{\inttd}{\int{d^2r}}
\newcommand{\jmag}{{j}_{\rm m}}
\newcommand{\jmagv}{\bm{j}_{\rm m}}
\newcommand{\jmagtd}{{j}_{\rm m}^{\rm (2d)}}
\newcommand{\jmagvtd}{\bm{j}_{\rm m}^{\rm (2d)}}
\newcommand{\kappatd}{\kappa^{\rm (2d)}}
\newcommand{\kappasw}{\kappa_{\rm sw}}
\newcommand{\Kperp}{K_\perp}
\newcommand{\sumrv}{\int\frac{d^3r}{a^3}}
\newcommand{\sumtd}{\int\frac{d^2r}{a^2}}
\newcommand{\setil}{\tilde{\se}}
\newcommand{\Simptil}{\tilde{S}^{\rm (i)}}
\newcommand{\ua}{u_{\rm c}}
\newcommand{\uT}{u_T}

%%%%%%%%%%%%%
\renewcommand{\Simpv}{{{\bm S}^{\rm (i)}}}
\renewcommand{\Simp}{{S^{\rm (i)}}}
\section{Thermal effects in spintronics}

Manipulation of magnetization and spin current without magnetic field is a key issue in spintronics technology. 
In the case of metallic ferromagnets, electric current has been proved to be highly useful for switching the magnetization and for driving magnetic domain walls by use of current-induced spin-transfer torques proposed by Berger \cite{Berger86,Berger96} and Slonczewski \cite{Slonczewski96}.
Metallic systems have been  studied extensively also for spin current manipulation such as for direct and inverse spin Hall effects \cite{Saitoh06,Kimura07}.
Besides metallic magnets, insulating ferrimagnet such as yttrium iron garnet (YIG) are also expected to be useful spintronics materials \cite{Kajiwara10} because of its weak spin relaxation effects as represented by small Gilbert damping parameter, $\alpha$.
Although use of electric current do not apply to insulators,  recent extensive studies revealed that manipulation of magnetization of YIG is possible by several different methods, such as using a temperature gradient \cite{Uchida08} and sound waves\cite{Uchida11}.
Thermal methods are of particular importance for realizing novel thermoelectric materials based on magnetic materials \cite{Hatami07,Kirihara12}, which are expected to be useful for devices with low energy consumption.

Recently, thermally-driven domain wall motion in a YIG film was observed optically \cite{Jiang13}.
The domain wall was found to move to a hotter side of the system, and the speed was about $2\times 10^{-4}$ m/s when a temperature gradient of $\nabla T=20$ K/mm is applied. 
The direction of the motion is counter-intuitive, but is consistent with the spin-transfer effect of spin waves (magnons) \cite{Hinzke11,Yan12}.
In fact, magnons excited around the wall are pushed by the temperature gradient to the colder side, and they transfer an spin angular momentum opposite to the local magnetization, resulting in a domain wall motion towards the hotter end. 
Thermodynamic viewpoint of domain wall motion was discussed by Wang \cite{Wang14} by evaluating numerically the entropy and free energy due to the magnons around the wall.
It was concluded that the wall motion to the hotter side is consistent with minimization of free energy when the entropy due to magnons are taken into account.
In a metallic ferromagnet, thermal motion of domain walls was observed in 1986 \cite{Jen_b86}.
The motion was towards the colder end, and the wall-entropy force proportional to $\nabla T$ was suggested as a possible mechanism \cite{Berger85,Jen_a86}, but the behavior would be explained simply by the spin-transfer torque induced by thermally-driven conduction electrons.

Domain wall under temperature gradient was numerically studied by solving the Landau-Lifshitz-Gilbert (LLG) equation for spins
in Ref. \cite{Hinzke11}, where the effect of temperature was modeled as a random magnetic field satisfying the fluctuation-dissipation theorem.
The results indicated that the behavior of the wall are essentially the same as the electric-current-driven case \cite{Thiaville05,TTKSNF06,TKS_PR08}. 
For instance, a sliding motion with a constant tilting angle occurs at low driving force or at large damping cases, while the tilting angle becomes time-dependent resulting in a screw-like motion when the driving force exceeds a threshold value corresponding to the Walker's break down field.
The LLG approach was employed also to study the motion of magnetic skyrmions \cite{Kong13}.
Skyrmions also turned out to move to the hotter region in the same manner as domain walls and the velocity was found to be inversely proportional to the Gilbert damping parameter, $\alpha$.
The numerical result was explained by the ``Brownian motion'' of the skyrmion driven by thermal random magnetic field and the spin-transfer effect from thermal spin waves. It was demonstrated that the spin wave spin-transfer effect contributing to the motion towards the hotter side is proportional to $\nabla T/\alpha$, while the velocity due to 
the Brownian motion, which pushes the skyrmions to the colder side, proportional to  $\alpha \nabla T$, and thus is negligibly small when $\alpha$ is small.
The study of Ref. \cite{Kong13} was extended to include dissipative torque arising from spin relaxation in Refs. \cite{Kovalev14,SeKwonKim15}.

\section{Luttinger's formulation of thermal transport}

Most of the theories on thermally-driven magnetic systems so far are either phenomenological \cite{Jen_a86,Kovalev12,Jiang13} or numerical \cite{Hinzke11,Kong13} and lack microscopic viewpoint.
The reason is obvious; the temperature gradient appears not to be straightforwardly integrated into a linear response theory or field-theoretical methods since its effect is not described by a microscopic interaction Hamiltonian.
This difficulty was removed in the case of electron transport phenomena by Luttinger already in 1964 \cite{Luttinger64}.
He introduced a fictitious scalar field called a \textquoteleft gravitational\textquoteright\ potential, $\Psi$,  which couples to the local energy density, ${\cal E}$, by an interaction Hamiltonian 
\begin{align}
  H_{\rm L} &= \intrv \Psi {\cal E}.
  \label{HTLuttinger}
\end{align}
It was argued that $\Psi$ satisfies $\nabla\Psi=\frac{\nabla T}{T}$ and that a linear response theory (Kubo formula) applies to the thermally-driven case by considering the correlation functions of the energy current density.
Many works have been carried out based on the \textquoteleft gravitational\textquoteright\ potential formalism on electron transport \cite{Smrcka77,Oji85,Cooper97,Catelani05,Michaeli09,Qin11,Schwiete14,EichDFT14}, Quantum dot \cite{Eich14}, magnon transport \cite{Matsumoto11b,Matsumoto11a,Matsumoto14} and thermal torques \cite{Kohno14}. 
At the same time, it has been noted that naive application of Kubo formula may result in wrong thermal coefficients \cite{Oji85,Smrcka77,Qin11,Kohno14}.

As alternative approaches for thermal transport, Landauer-like approach by connecting locally equilibrium systems at different temperatures is possible \cite{Butcher90,Takezoe10}.
It was shown in Ref. \cite{Butcher90} that the obtained transport coefficients have no unphysical divergence, since only the electrons at the Fermi energy contribute to them.
Quantum kinetic equation approach was employed in Refs. \cite{Catelani05,Michaeli09} to study electron correlation effects.
Still, a fully quantum mechanical description based on a linear response theory has wider applications and is highly useful.

Prescriptions to extract correct results based on the \textquoteleft gravitational\textquoteright\  potential formalism were given by some works \cite{Smrcka77,Qin11,Kohno14}.
In the case of thermally-induced electron transport, Smrcka and Streda carried out the calculation of thermal Hall effect in the presence of an applied magnetic field,  based on the Luttinger's prescription \cite{Smrcka77,Oji85}.
They rewrote the Luttinger's interaction, $\Psi {\cal E}$, by use of integral by parts as  $-{\cal E}\rv\cdot\nabla \Psi$ and carried out a perturbative expansion with respect to a physical force proportional to $\nabla \Psi=\nabla T/T$.
The analysis, however, is based on a interaction proportional to a unbounded operator, $\rv$, and thus the results may not be always convincing.
Recent works on thermal magnons are based on the same treatment \cite{Matsumoto14}.
Later, Qin et. al. pointed out that a naive application of Luttinger's approach for thermal Hall effect leads to a unphysical divergence at $T\ra0$, and that this divergence arises from an equilibrium rotational electron current induced when the time-reversal invariance is broken by the magnetic field \cite{Qin11}. 
They showed that the correct non-equilibrium response is obtained if one subtract the equilibrium current before applying the \textquoteleft gravitational\textquoteright\  potential.

The case of thermally-induced torque in ferromagnetic metals was studied in detail by Kohno et. al. \cite{Kohno14}.
They calculated the non-equilibrium torque on the magnetization generated by the conduction electron when a temperature gradient is applied and found that unphysical divergence arises if Luttinger's approach is straightforwardly applied. They demonstrated that the divergence arises from the equilibrium torque describing the exchange interaction between the magnetization and that this equilibrium contribution needs to be removed when discussing non-equilibrium torque.

\section{Vector potential formulation}

It was recently pointed out in Ref. \cite{Tatara15} that those problems of Luttinger's formalism arise from the fact that the \textquoteleft gravitational\textquoteright\  potential couples to the total energy density, modifying the equilibrium properties in addition to inducing non-equilibrium response.
It was discussed that   the role of diamagnetic current, which is essential for removing unphysical equilibrium (non-dissipative) contribution from transport coefficients \cite{Heuser74,Qin11}, is not seen straightforwardly in the scalar potential formalism.
Instead of a scalar potential formalism, a vector potential formalism to describe thermally-induced transport was developed in Refs. \cite{Shitade14,Tatara15}.
In Ref. \cite{Shitade14}, Shitade demonstrated using the analogy of general relativity that if an invariance under time translation is imposed locally, a vector potential arises from Luttinger's scalar potential and they are described by a gauge invariant theory.
He applied his model to describe thermal Hall effect of non-interacting electrons and showed that the results satisfy the Wiedemann-Franz law.
The origin of the invariance under local time translation was not argued. 
In Ref. \cite{Tatara15}, the vector potential was introduced by rewriting the Luttinger's Hamiltonian for the \textquoteleft gravitational\textquoteright\  potential using the conservation law of the energy in the limit of static (dc) limit of the temperature gradient.
The vector potential interaction was also argued there in the context of entropy force. 
Since the thermal vector potential couples to energy current, it only generate excitations and does not alter equilibrium contributions. 
The  vector potential representation thus guarantees a straightforward linear response calculations for thermal dynamics on the equal footing as the electric field-driven case.
It was indeed shown that unphysical equilibrium contributions are canceled automatically by \textquoteleft diamagnetic\textquoteright\  currents associated with the vector potential. 
A possibility of vector potential description was briefly mentioned in Ref. \cite{Miura12} in discussing magnon-drag thermoelectric effects.

The vector potential form of the interaction Hamiltonian describing the thermal effect is 
\begin{align}
 \HAT
  &\equiv -\intrv \jv_{\cal E}(\rv,t)\cdot \Av_{T}(t)
  \label{HATdef}
\end{align}
where $ \Av_{T}(t)$ 
is the thermal vector potential. It satisfies 
\begin{align}
   \partial_t \Av_{T}(\rv,t) 
    = \nabla\Psi(\rv,t)=\frac{\nabla T}{T},
\end{align}
where $\Psi$ is the Luttinger's \textquoteleft gravitational\textquoteright potential. 
The interaction (\ref{HATdef}) was used to describe thermally-induced longitudinal transport and Hall effect of non-interacting electrons \cite{Shitade14,Tatara15}.
In the present paper, we apply the formalism to study thermally-induced magnetization dynamics.

\section{Energy current of localized spins}

To study thermal transport based on Eq. (\ref{HATdef}), we need to derive the expression for the energy current density.
Following Ref. \cite{Tatara15}, we carry out the derivation quantum mechanically by use of energy conservation law,
\begin{align}
\dot{\cal E}+\nabla\cdot\jv_{\cal E}=0,
\end{align}
where ${\cal E}$ is the energy density.
We consider a case of a ferromagnet with an easy-axis and a hard-axis magnetic anisotropy energies. 
The Hamiltonian is $H \equiv  \sumrv {\cal E}$, where 
\begin{align}
{\cal E} &= \frac{J}{2}  (\nabla\Sv)^2 - \frac{K}{2} (S_z)^2 + \frac{\Kperp}{2} (S_y)^2
   ,   \label{HamiltonianDW}
\end{align}
where $J$, $K(>0)$ and $\Kperp(\geq0)$ represent the energy of the exchange interaction, the easy-axis anisotropy and the hard-axis anisotropy, respectively, and $\al$ is the atomic lattice constant.
The energy current density in this case is obtained as (see Sec. \ref{SECJEspin} for derivation)
\begin{align}
  \jv_{{\cal E},i} 
  & = 
   - \frac{J}{a^3}  \nabla_i\Sv\cdot\dot{\Sv} .\label{JEspin}
\end{align}
Thermally-driven dynamics of the spin structure is calculated below as a linear response to the interaction $\HAT$, Eq. (\ref{HATdef}), with the energy current density of Eq. (\ref{JEspin}).

\section{Separation of variables \label{SECseparation}}

To carry out the calculation of thermally-driven magnetization dynamics, we separate a collective degrees of freedom describing a classical magnetization structure and fluctuation, magnons or spin waves, around the structure.
The direction of the localized spins for the classical solution is represented by use of polar angles $\theta(\rv,t)$ and $\phi(\rv,t)$. 
The magnon excitation around the structure is then represented by use of the Holstein-Primakov boson defined with respect to the local quantization axis along $(\theta(\rv,t),\phi(\rv,t))$. 
The localized spin vector, $\Sv$, is thus represented as
\begin{align}
  \Sv &= U(\rv,t) \widetilde{\Sv} \equiv  U(\rv,t) (S\hat{\zv}+\delta \sv),
  \label{Sdef}
\end{align}
where $\hat{\zv}$ is the unit vector along the $z$ axis, $U$ is a $3\times3$ unitary matrix describing a rotation of a vector $\hat{\zv}$ to the direction $(\theta,\phi)$,  
$\widetilde{\Sv}\equiv S\hat{\zv}+\delta \sv$, and $\delta \sv$ represents the fluctuation.
The unitary matrix is chosen as \cite{Thiele73}
\begin{align}
  U &= \lt( \begin{array}{ccc}
         \cos\theta\cos\phi & -\sin\phi & \sin\theta\cos\phi  \\
         \cos\theta\sin\phi &  \cos\phi & \sin\theta\sin\phi  \\
         -\sin\theta & 0 & \cos \theta 
       \end{array}  \rt),
\end{align}
and the fluctuation is represented in terms of annihilation and creation operators for magnon (the Holstein-Primakov boson), $\boson$ and $\boson^\dagger$, as 
\cite{Kittel63}
\begin{align}
  \delta \sv &= \lt( \begin{array}{c}
         \gamma (\boson^\dagger +\boson) \\
         i\gamma(\boson^\dagger-\boson) \\
         - \boson^\dagger \boson 
       \end{array}  \rt),\label{HPboson}
\end{align}
where $\gamma\equiv \sqrt{\frac{S}{2}}$.
We neglect the terms third- and higher-order in boson operators.

The unitary transformation modifies derivatives of spin as ($\mu=t,x,y,z$)
\begin{align}
  \partial_\mu\Sv=U(\partial_\mu+i\AU) \widetilde{\Sv},
\end{align}
where 
\begin{align}
 {\AU}_{,\mu}   & \equiv -i U^{-1} \nabla_\mu U,
\end{align}
is a spin gauge field represented by a $3\times3$ matrix.
Explicitly, the spin gauge field reads 
\begin{align}
 {\AU}_{,\mu}   & = -i \nabla_\mu\theta 
        \lt( \begin{array}{ccc}
         0 & 0 & 1 \\
         0 & 0 & 0  \\
         -1 & 0 & 0
       \end{array}  \rt)
          -i\nabla_\mu\phi 
           \lt( \begin{array}{ccc}
         0 & -\cos\theta & 0 \\
         \cos\theta & 0 & \sin\theta  \\
         0 & -\sin\theta & 0
       \end{array}  \rt).
\end{align}
The energy current density, Eq. (\ref{JEspin}), is then 
\begin{align}
 j_{{\cal E},i}
  &= - \frac{J}{ \al^3} \widetilde{\Sv}(\stackrel{\leftarrow}{\partial_t}-i{\AU}_{,t})(\nabla_i+i{\AU}_{,i}) \widetilde{\Sv}
   \nnr
  &= - \frac{J}{ \al^3} \lt[ 
  S^2 ( (\nabla_i\theta)\dot{\theta}+\sin^2\theta(\nabla_i\phi)\dot{\phi} )
   \lt(1-\frac{1}{S}\boson^\dagger \boson \rt) \rt. \nnr
 & +S\lt(\dot{\boson^\dagger}(\nabla_i \boson)+(\nabla_i \boson^\dagger)\dot{\boson} \rt)
  \nnr
 & +S\lt[ -i\cos\theta[ \dot{\phi} (\dot{\boson^\dagger}\nablalr \boson) +(\nabla_i \phi)(\boson^\dagger\stackrel{\leftrightarrow}{\partial_t}{\boson})  ] \rt. \nnr
  & \lt. +2\cos^2\theta(\nabla_i \phi)\dot{\phi} \boson^\dagger \boson \rt] \nnr
 &+ 
  S\gamma\lt[ -i\cos\theta[\dot{\theta}(\nabla_i \phi) + (\nabla_i\theta) \dot{\phi}](\boson^\dagger-\boson) 
  \rt. \nnr &  \lt.
  +(\dot{\theta}+i\sin\theta \dot{\phi}) (\nabla_i \boson^\dagger) 
  +(\dot{\theta}-i\sin\theta \dot{\phi}) (\nabla_i \boson) 
  \rt. \nnr & \lt. \lt.
  +(\nabla_i{\theta}+i\sin\theta \nabla_i{\phi}) (\dot{\boson^\dagger}) 
  +(\nabla_i{\theta}-i\sin\theta \nabla_i{\phi}) (\dot \boson)  \rt]\rt].
\label{Jefull}
\end{align}
The first term,
\begin{align}
 j_{{\cal E},i}^{\rm (s)}
  &\equiv - \frac{J}{\al^3}
  S^2 ( (\nabla_i\theta)\dot{\theta}+\sin^2\theta(\nabla_i\phi)\dot{\phi} )
   \lt(1-\frac{1}{S}\boson^\dagger \boson \rt),\label{JEsdef}
\end{align}
is the energy current carried by the magnetization structure, and 
\begin{align}
 j_{{\cal E},i}^{\rm (m)}
  &\equiv 
 - \frac{J S}{\al^3} \lt[ \dot{\boson^\dagger}(\nabla_i \boson)+(\nabla_i \boson^\dagger)\dot{\boson} \rt]
, \label{JEswdef}
\end{align}
is the contribution of magnons.
We see that the temperature gradient appears to act on the magnetization structure as well as on the  spin waves.
We need to be  careful, however, since Eq. (\ref{JEsdef}) may contain a steady-state contribution not contributing to the thermally-excited response (see Sec. \ref{SECDW}). 
The mixed contributions in Eq. (\ref{Jefull}) where the energy current is carried by both the spin texture and the magnon turn out to be second order in the temperature gradient and are thus neglected.
The last contributions linear in the magnon operators are neglected similarly.
It is thus sufficient to consider the two energy currents, Eqs. (\ref{JEsdef})(\ref{JEswdef}) in the following analysis of linear response regime.

\section{Magnon spin-transfer effect}

The unitary transformation gives rise to a gauge field, $\AUv$, due to the exchange interaction according to 
\begin{align}
  (\nabla\Sv)^2 %&=  |\nabla \widetilde{\Sv} +i\AUv\widetilde{\Sv}|^2 
    = (\nabla\widetilde{\Sv})^2  -i \widetilde{\Sv}^\dagger (\AUv \cdot \nablalr) \widetilde{\Sv} + \widetilde{\Sv}^\dagger (\AU)^2\widetilde{\Sv},
\end{align}
where $\nablalr_\mu \equiv \stackrel{\rightarrow}{\nabla_\mu}-\stackrel{\leftarrow}{\nabla_\mu}$ acts on field operators on the both sides but not on $\AUv$, i.e., 
$\widetilde{\Sv}^\dagger (\AUv \cdot \nablalr) \widetilde{\Sv}=\sum_{i} {\AU}_{,i}[\widetilde{\Sv}^\dagger (\nabla_i \widetilde{\Sv})-\ (\nabla_i \widetilde{\Sv}^\dagger) \widetilde{\Sv}]$.
The exchange interaction term is thus 
\begin{align}
  \frac{J}{2}\sumrv (\nabla \Sv)^2 &= \frac{J}{2}\sumrv (\nabla \widetilde{\Sv})^2 + \Hdwsw ,
\end{align}
where 
\begin{align}
  \Hdwsw & \equiv -i\frac{J}{2} \sumrv \widetilde{\Sv}^\dagger (\AUv \cdot \nablalr) \widetilde{\Sv}
      \nnr
  &= 4JS\intrv \Aadv\cdot\jmagv^{(0)},
 \label{Hdwsw}
\end{align}
represents the interaction between the spin structure and the magnons (subscript st denotes spin-transfer), and contributions second order in $\AU$ are neglected. 
Here
$ \Aadv \equiv \frac{1}{2}\cos\theta\nabla\phi$ is the adiabatic component of spin gauge field \cite{TKS_PR08} and 
\begin{align}
  \jmagv^{(0)} \equiv -\frac{i}{2\al^3}(\boson^\dagger \nablalr \boson),\label{jmagpara}
\end{align}
is the magnon current density. 
As seen in (\ref{Hdwsw}), the magnon current couples to the adiabatic spin gauge field in the same manner as the case of the electric current in metals in the adiabatic regime \cite{TK04,TKS_PR08}.
The magnon current therefore induces the spin-transfer torque with the angular momentum transfered per area and per unit time is $4JS\jmag$ (instead of $\frac{\hbar}{e} P j$ in the case of electric current ($P$ is the spin polarization of the current and $e$ is the electron charge)).

There is another contribution to the interaction between the magnetization structure and magnons, namely, the one arising from $\HAT$ (Eq.(\ref{HATdef})).
Using Eq. (\ref{Jefull}), the spin-transfer terms in  $\HAT$ is 
\begin{align}
  \HstAT  & \equiv i\frac{J}{\al^3}A_{T,i} \intrv 
   [ (\partial_t\widetilde{\Sv}) {\AU}_{,i} \widetilde{\Sv}
   +(\nabla_i\widetilde{\Sv}) {\AU}_{,t} \widetilde{\Sv}].
\end{align}
Keeping only the contributions second order in the magnon operators, it reads 
\begin{align}
  \HstAT  & = 4JS\intrv \Aadv\cdot\jmagv^{\rm (d)},
\end{align}
where 
\begin{align}
  \jmagv^{\rm (d)} 
    \equiv -\frac{i}{2\al^3}\Av_T(\boson^\dagger \stackrel{\leftrightarrow}{\partial_t} \boson),
    \label{jmagdia}
\end{align}
is \textquoteleft diamagnetic\textquoteright\ current of magnon.
The total magnon spin-transfer effect is thus described by
\begin{align}
  \Hst  & = 4JS\intrv \Aadv\cdot\jmagv,
\end{align}
where $ \jmagv\equiv \jmagv^{(0)}  + \jmagv^{\rm (d)} $.

\section{Domain wall \label{SECDW}}

Let us describe a domain wall based on the Hamiltonian (\ref{HamiltonianDW}).
Considering a thin wire, the magnetization structure is treated as one-dimensional, i.e., changing only in the wire direction, which we choose as $z$ axis.
The classical solution is determined by the equation of motion as
\begin{align}
\cos\theta &= \tanh \frac{z-X(t)}{\lambda}, & 
\sin\theta &= \frac{1}{ \cosh \frac{z-X(t)}{\lambda}} ,
\end{align}
and $\nabla_z\phi(t)=0$, where $\lambda\equiv \sqrt{\frac{J}{K}}$ is the thickness of the domain wall.
We then have 
\begin{align}
\nabla_z \Sv &= -\frac{S}{\lambda}\ev_\theta,
\end{align}
where $\ev_{\theta}\equiv(\cos\theta\cos\phi,\cos\theta\sin\phi,-\sin\theta)$.

We first show that the energy current of the wall given by Eq. (\ref{JEsdef}) is an equilibrium contribution, which should not be taken into account.
Let us first see what happens if Eq. (\ref{JEsdef}) is naively applied.
Equation (\ref{JEsdef}) for domain wall reads 
(neglecting the corrections of the order of $S^{-1}$ by magnons) 
\begin{align}
 j_{{\cal E},i}^{\rm (s)}
  & =  \delta_{i,z} \frac{JS^2 }{\lambda^2 \al^3} \dot{X} \frac{1}{\cosh^2 \frac{z-X(t)}{\lambda}}.
  \label{JEDW}
\end{align}
The coupling between the domain wall and temperature gradient, Eq. (\ref{HATdef}), then becomes 
\begin{align}
 \HAT
   = \frac{2AJS^2}{\al^3\lambda}\frac{\nabla_z T}{T}  \int^t_0 dt'\dot{X}(t')
   =  \Nw \frac{JS^2 }{\lambda^2} \frac{\nabla_z T}{T} X,
\end{align}
where $\Nw\equiv 2\lambda A/a^3$ is the number of spins in the wall ($A$ is the cross sectional area of the wire).
One may be tempted to conclude that the thermal force on the wall is 
\begin{align}
 F_T &\equiv -\frac{\delta \HAT}{\delta X}   
      =  - \Nw \frac{JS^2 }{2\hbar \lambda^2} \frac{\nabla_z T}{T}.
\end{align}
As we see, this force diverges at $T\ra0$ if $\nabla T$ is fixed, which is unphysical. 
The problem arose from the energy current density of Eq. (\ref{JEDW}).  
Equation (\ref{JEDW}) indicates that the energy current arises proportional to the wall speed $\dot{X}$. This energy current is, however, a steady-state current, which does not contribute to the entropy change, since the energy carried by the domain wall is not converted to a heat unless the wall is annihilated.
To consider the standard experimental situation where the wall is semi macroscopic and does not annihilate, one needs to thus subtract the steady-state contribution from  Eq. (\ref{JEDW}), resulting in vanishing energy current for the wall. 
This treatment is consistent with the argument in Ref. \cite{Wang14}
The direct thermal force would become physical, if we consider an ensemble of domain walls which are allowed to annihilate and to be created (see Sec. \ref{SECvortex}). 

The thermal effect for a domain wall therefore arises only from the magnon interaction, $\Hdwsw$ (Eq. (\ref{Hdwsw})). 
Noting that 
$\frac{\delta {\Aad}_{,i}}{\delta \phi} = \frac{1}{2}\sin\theta\nabla_i\theta$, 
the force and torque induced by the magnon interaction read
\begin{align}
F_{\rm m} & \equiv - \frac{\delta \Hdwsw}{\delta X}  = 0 \nnr
\torque_{\rm m} &\equiv \frac{\delta \Hdwsw}{\delta \phi} = -\frac{ 2JS}{\lambda}j_{{\rm m},z}\intrv \frac{1}{\cosh^2\frac{z}{\lambda}} = - \frac{2\Nw JS\al^3}{\lambda}j_{{\rm m},z}.
\label{forcetorquemagnon}
\end{align}
In Eq. (\ref{forcetorquemagnon}), the force from the magnon disappears since we consider here a static one-dimensional domain wall, in which case the magnons are in the perfect adiabatic limit \cite{YanMagnon11}; In other words, magnon modes are orthogonal to the collective modes of a static one-dimensional domain wall \cite{TKS_PR08}.
The force from the magnons may arise when the wall is dynamic \cite{LeMaho09} or higher-dimensional and in the presence of dipolar interaction \cite{Yan13}.
Here we introduce phenomenologically a thermal force by use of a dimensionless parameter $\beta_T$ as 
$F_\beta\equiv -\beta_T \frac{\Nw S}{\lambda^2}\kb a^2 \nabla_z T$.
Including this force and the thermal magnon effects, the equation of motion for the wall is 
\begin{align}
 \dot{\phi}+\alpha\frac{\dot{X}}{\lambda}  &= - \beta_T \frac{\kb a^2 }{\hbar \lambda}\nabla_z T \nnr
 \dot{X}-\alpha\lambda\dot{\phi} &= \frac{\Kperp\lambda S}{2\hbar}\sin 2\phi + \frac{\lambda}{\hbar \Nw S}\torque_{\rm m} \nnr 
& = \frac{\Kperp\lambda S}{2\hbar}\sin 2\phi - \frac{2J\al^3}{\hbar} j_{{\rm m},z}.
 \label{DWeq}
\end{align}

The magnitude of the magnon current is evaluated by use of a linear response theory with respect to $\HAT$ in Sec. \ref{SEClinear}.

\section{Vortex \label{SECvortex}}

Let us next consider a case of a single vortex which appears in a two-dimensional sub micron size disks. 
A Hamiltonian describing the vortex is
\begin{align}
H &= \sumtd\lt[\frac{J}{2}  (\nabla\Sv)^2 - \frac{K}{2} (S_z)^2   \rt]
   ,   \label{Hamiltonianvortex}
\end{align}
and the vortex structure with vortex number of unity is approximately represented as \cite{Shibata06}
\begin{align}
  \Sv_{\rm v}(\rv)=S\lt(\cos\lt(\varphi+\frac{\pi}{2}c\rt)\ev_x + \sin\lt(\varphi+\frac{\pi}{2}c\rt)\ev_y\rt),
\end{align}
where $\tan\varphi=\frac{y}{x}$ and $c$ is an integer representing the chirality.
Using the collective coordinates, $X(t)$ and $Y(t)$, representing the center of the vortex in two-dimensions, the spin structure is $\Sv(\rv,t)=\Sv_{\rm v}(\rv-\Xv(t))$, where $\Xv(t)\equiv (X(t),Y(t))$.
Similarly to the domain wall case, we consider the case where vortex cannot annihilate;
All the thermal effect then arises form magnons. 
The spin-transfer torque due to magnons, Eq. (\ref{Hdwsw}), is written when magnon current is uniform as 
\begin{align}
  \Hdwsw &= \frac{2J}{S^2}\inttd \Sv\cdot[(\jmagvtd\cdot\nabla)\Sv\times\delta\Sv],
\end{align}
where $\delta \Sv$ is the change of $\Sv$ when the origin of $\Xv$ is shifted by an amount $\delta\Xv=(\delta X,\delta Y)$. 
(This representation is convenient to focus on physical contribution in topological quantities for spin  \cite{Auerbach94}. ) 
The magnon current density $\jmagvtd$ is the two-dimensional one ($\jmagvtd=\jmagv L_z$ where $L_z$ is the thickness of the system).
The force due to the spin-transfer is therefore
\begin{align}
  F_{{\rm st},i} & \equiv -\frac{\delta \Hdwsw}{\delta X_i} =-\frac{2J}{S^2}\inttd \Sv\cdot[\nabla_i\Sv\times\nabla_j\Sv] {j}_{{\rm m},j}^{\rm (2d)}.
\end{align}
This force is characterized by the topological number density of the vortex,
\begin{align}
  G &= \frac{\hbar}{S^2}\sumtd \Sv\cdot[\nabla_x\Sv\times\nabla_y\Sv] =\frac{2\pi \hbar S}{\al^2},
\end{align}
as 
\begin{align}
  \Fv_{{\rm st}} & = \frac{2J\al^2}{\hbar} \Gv\times \jmagvtd,
\end{align}
where $\Gv\equiv G\hat{\zv}$.
The equation of motion of a vortex is then \cite{Thiele73,Shibata06}
\begin{align}
   -\Gv\times \dot{\Xv} +\alpha D\dot{\Xv} = \Fv_{\rm st},
\end{align}
where 
$D\simeq \frac{\hbar S}{\al^2}\int d^2r \sin^2\theta(\nabla\phi)^2$ is a form factor for the damping.
The equation is written as
\begin{align}
   \dot{\Xv}+\frac{2J\al^2}{\hbar}\jmagvtd = -\frac{\alpha D}{G^2}\Gv\times\dot{\Xv} .
\end{align}
The velocity is thus  
\begin{align}
   \dot{\Xv}= - \frac{1}{1+\lt(\frac{\alpha D}{G}\rt)^2}\lt(1+i\sigma_y \frac{\alpha D}{G}\rt)   \frac{2J\al^2}{\hbar} \jmagvtd,
\end{align}
where $\sigma_y$ is the $y$-component of the Pauli matrix.
We consider the case of temperature gradient along the $x$-direction. Since the magnon current is along $\nabla T$, the vortex velocity is 
\begin{align}
   \lt(\begin{array}{c} \dot{X} \\ \dot{Y} \end{array} \rt)
  = \frac{1}{1+\lt(\frac{\alpha D}{G}\rt)^2} \frac{2J\al^2}{\hbar} \jmagtd 
    \lt(\begin{array}{c} 
   -1  \\ 
    \frac{\alpha D}{G}  \end{array} \rt).\label{Vortexvelocity}
\end{align}

In the study by a numerical simulation and the Focker-Planck equation in Ref. \cite{Kong13} carried out for a skyrmion, there was a direct force term proportional to $\nabla_x T$.
(A single skyrmion and a vortex are described the the same equation called the Thiele equation \cite{Thiele73} and thus have the same response to applied forces.) 
Such direct force arises in our formalism by taking account of annihilation of vortices and by regarding the energy current, Eq. (\ref{JEsdef}), as physical one contributing to the entropy production.
The energy current density in this case becomes 
$\jv_{\cal E}=\gamma_{\rm v}\dot{\Xv}$ ($\gamma_{\rm v}$ is a temperature-dependent coefficient), and  the direct thermal force becomes 
\begin{align}
  \Fv_T &= - \frac{\delta \HAT}{\delta \Xv} = \frac{\gamma_{\rm v}}{T} \nabla T.
\end{align}
Since the annihilation of vortices requires a finite excitation energy, $\gamma_{\rm v}$ vanishes quickly at $T=0$, and the force would  vanish at $T=0$.
It would be interesting to study whether the direct thermal force  exists or not experimentally.

\section{Linear response theory of magnon current  \label{SEClinear}}

We calculate the magnon current induced by the temperature gradient within the linear response theory.
We carry out the calculation by use of the non-equilibrium Green's function. 
(Standard Kubo formula calculation is also applicable.)
The interaction $\HAT$ in the Fourier representation of the magnon operators 
($\boson(\rv,t)= \sqrt{\frac{\al^3}{V}}\sumom\sumqv e^{i(\qv\cdot\rv-\omega t)} \boson_{\qv,\omega}$) is
\begin{align}
 \HAT & 
  = -2JS \sumom\sumOm \frac{1}{V}\sumqv e^{i\Omega t} q_i \omega A_{T,i}(-\Omega) \boson^\dagger_{\qv,\omega+\frac{\Omega}{2}}\boson_{\qv,\omega-\frac{\Omega}{2}},
\end{align}
where $\Omega$ is an infinitesimal external angular frequency and  $V$ is the system volume.
The \textquoteleft paramagnetic\textquoteright\ and \textquoteleft diamagnetic\textquoteright\ magnon currents, Eq. (\ref{jmagpara}) and (\ref{jmagdia}), respectively, are
\begin{align}
  j_{{\rm m},i} ^{(0)}
    &= i \sumom\int\frac{d\Omega}{2\pi} \frac{1}{V} \sum_{\qv} e^{i\Omega t} 
    {q_i} G^<_{\qv,\omega-\frac{\Omega}{2},\qv,\omega+\frac{\Omega}{2}} \nnr
  j_{{\rm m},i} ^{\rm (d)}
    &= -i \sumom\int\frac{d\Omega}{2\pi} \frac{1}{V} \sum_{\qv} e^{i\Omega t} 
    \omega A_{T,i}(-\Omega) G^<_{\qv,\omega-\frac{\Omega}{2},\qv,\omega+\frac{\Omega}{2}} ,
    \label{jmagdef}
\end{align}
where $ G^<_{\qv,\omega,\qv',\omega'}\equiv -i\average{ \boson^\dagger_{\qv'\omega'} \boson_{\qv\omega} } $ is the lesser component of the non-equilibrium Green's function of magnons
\cite{Kadanoff94}.
The \textquoteleft paramagnetic\textquoteright\ contribution is calculated at the linear order in $\HAT$ as
\begin{align}
 j_{{\rm m},i} ^{(0)}
  &= -i \frac{2JS}{V} \sumom\int\frac{d\Omega}{2\pi}  \sum_{\qv} q_i q_j \omega \lt[g_{\qv,\omega-\frac{\Omega}{2}} g_{\qv,\omega+\frac{\Omega}{2}}\rt]^<.
\end{align}
The lessor Green function is written in terms of retarded and advanced Green's functions as \cite{Haug07}
\begin{align}
\biggl[ & g_{\qv,\omega-\frac{\Omega}{2}} g_{\qv,\omega+\frac{\Omega}{2}}\biggr]^< \nnr
  &= \lt(n\lt(\omega+\frac{\Omega}{2}\rt)-n\lt(\omega-\frac{\Omega}{2}\rt)\rt) \gr_{\qv,\omega-\frac{\Omega}{2}} \ga_{\qv,\omega+\frac{\Omega}{2}} \nnr
 &
     -n\lt(\omega+\frac{\Omega}{2}\rt) \gr_{\qv,\omega-\frac{\Omega}{2}} \gr_{\qv,\omega+\frac{\Omega}{2}}
      + n\lt(\omega-\frac{\Omega}{2}\rt) \ga_{\qv,\omega-\frac{\Omega}{2}} \ga_{\qv,\omega+\frac{\Omega}{2}}     \nnr
 &\simeq 
n\lt(\omega\rt) [ (\ga_{\qv,\omega})^2 -  (\gr_{\qv,\omega})^2 ] 
-\frac{\Omega}{2} n'\lt(\omega\rt) (\ga_{\qv,\omega} -\gr_{\qv,\omega})^2  
 ,
\end{align}
where $n(\omega)\equiv [e^{\beta\omega}-1]^{-1}$ is the Bose distribution function, $n'(\omega)\equiv \frac{dn}{d\omega}$, $\beta\equiv 1/(\kb T)$ ($\kb$ is the Boltzmann constant) and we have neglected contribution of the order of $\Omega^2$.
The retarded and advanced Green's function are 
\begin{align}
\gr_{\qv\omega} & = \frac{1}{\omega-\omega_q+i\alpha\omega}, 
\end{align}
and $\ga_{\qv\omega}=(\gr_{\qv\omega})^*$, 
where $\omega_q$ is the angular frequency of magnon with a wave vector $\qv$ and the effect of Gilbert damping is included as an imaginary part.
The \textquoteleft paramagnetic\textquoteright\ magnon current is thus
\begin{align}
  j_{{\rm m},i} ^{(0)}
    &= JS \sumom\int\frac{d\Omega}{2\pi} \frac{1}{V}\sum_{\qv} q_i q_j 
     (i\Omega) A_{T,j}(-\Omega)\nnr
 & \times  \omega n'(\omega)\lt( \ga_{\qv\omega} -\gr_{\qv\omega} \rt)^2 \nnr
    & + i\sumom\int\frac{d\Omega}{2\pi} \frac{1}{V}\sum_{\qv}  
      A_{T,i}(-\Omega) \omega n(\omega)\lt( \ga_{\qv\omega} -\gr_{\qv\omega}\rt) ,
\end{align}
where we used $2JS q_j (\ga_{\qv\omega})^2=\partial_{q_j} \ga_{\qv\omega}$.
The \textquoteleft diamagnetic\textquoteright\ magnon current is similarly calculated as 
\begin{align}
  \jmagv^{\rm (d)} 
   = -i\sumom\int\frac{d\Omega}{2\pi} \frac{1}{V}\sum_{\qv}   A_{T,i}(-\Omega) 
  \omega n(\omega)\lt( \ga_{\qv\omega} -\gr_{\qv\omega}\rt).
\end{align}
We see that the  \textquoteleft diamagnetic\textquoteright\ current cancels an equilibrium contribution of the  \textquoteleft paramagnetic\textquoteright\ current, resulting in (assuming rotational symmetry and using 
$\sumOm %e^{-i\Omega t}
(-i\Omega) A_{T,i}(-\Omega)= \dot{\Av}_{T}=\frac{\nabla T}{T}$) 
\begin{align}
  j_{{\rm m},i} 
  &= -\kappa \nabla_i T,\label{jkappadef}
\end{align}
where the coefficient $\kappa$ is 
\begin{align}
  \kappa 
   &=  \frac{JS}{3\hbar} \frac{1}{T} \sumom\frac{1}{V}\sum_{\qv}   q^2 \omega 
   n'(\omega)\lt( \gr_{\qv\omega} -\ga_{\qv\omega} \rt)^2 
 \nnr
   &= - \frac{4JS}{3\hbar} \alpha^2   \frac{1}{T} \frac{1}{V} \sum_{\qv} q^2 \sumom 
    n'(\omega) \frac{\omega^3}{[(\omega-\omega_q)^2+(\alpha\omega)^2]^2}.
\end{align}

To evaulate the summation over $\qv$, we consider the case of uniaxial anisotropy ($\Kp\ll K$) for simplicity . The magnon energy  is then $\hbar\omega_q= JSq^2+\gap$, where $\gap=KS$ is the gap of spin wave. 
The summation over $\qv$ is carried out in three dimensions.
Defining $\epsilon\equiv JSq^2+\gap$, we obtain 
\begin{align}
\frac{1}{V} & \sum_{\qv} q^2 \frac{1}{[(\omega-\omega_q)^2+(\alpha\omega)^2]^2}  \nnr
   &= \frac{1}{4\pi^2 (JS)^{5/2}}\int_{\gap}^\infty d\epsilon (\epsilon-\gap)^{3/2} 
      \frac{1}{[(\epsilon-\omega)^2+(\alpha\omega)^2]^2}  \nnr
   &=
       \frac{1}{2\pi (JS)^{5/2} \alpha^3}\theta(\omega-\gap) \frac{(\omega-\gap)^{3/2}}{\omega^3}
      ,
\end{align}
where $\theta(x)$ is step function and we used an approximation (using $\alpha\ll1$)
\begin{align}
  \frac{1}{[(\epsilon-\omega)^2+(\alpha\omega)^2]^2} 
    & \simeq \frac{2\pi}{(\alpha\omega)^3}\delta(\epsilon-\omega).
\end{align}
The result is thus 
\begin{align}
  \kappa 
   &= - \frac{1}{3\pi^2 (JS)^{3/2}} \frac{1}{\alpha T} \int_{\gap}^\infty d\omega (\omega-\gap)^{3/2}\frac{dn(\omega)}{d\omega}   \nnr
   &=  \frac{1}{2\pi^2 (JS)^{3/2}} \frac{1}{\alpha } \kb  (\kb T)^{1/2} F(\beta\gap),  
   \label{kappares3d}
\end{align}
where 
\begin{align}
   F(\beta\gap) 
     & \equiv \int_{\beta\gap}^\infty dx  \frac{(x-\beta\gap)^{1/2}}{e^{x}-1}.
\end{align}
At high temperatures, $\beta\gap \ll 1$, $F=2.315$. 
We see that the thermal transport coefficients are proportional to the damping time constant, $\alpha^{-1}$, in the same manner as in the electric transport coefficients in metals are proportional to the elastic lifetime, $\tau$. 
The result, $\kappa\propto \sqrt{T}/\alpha$, agrees with a semi-classical calculation in the supplement of Ref. \cite{Jiang13}.

\section{Domain wall solution  \label{SECdwmotion}}

Using the result of magnon current, Eq. (\ref{jkappadef}) with Eq. (\ref{kappares3d}), 
the equation of motion for a domain wall, Eq. (\ref{DWeq}), reads
\begin{align}
 \dot{\phi}+\alpha\frac{\dot{X}}{\lambda}  &=  \beta_T \frac{\uT}{\lambda} \label{DWeq1} \\
 \dot{X}-\alpha\lambda\dot{\phi} &= \vc\sin 2\phi  
   - P_T \uT ,
 \label{DWeq2}
\end{align}
where
\begin{align}
 \vc &\equiv \frac{\Kperp\lambda S}{2\hbar} ,
\end{align}
 and 
\begin{align}
 \uT &\equiv - \frac{\kb a^2}{\hbar}{\nabla_z T} 
\end{align}
represents the scale of the velocity induced by the temperature gradient (with positive direction chosen as the direction from the high- to the low-temperature region).
The spin-polarization coefficient in the thermal magnon spin-transfer is temperature-dependent as 
\begin{align}
P_T & \equiv \frac{2J\kappa}{\kb\al^2} 
= \frac{F}{\pi^2 \sqrt{S}} \frac{1}{\alpha}\sqrt{{\kb T}\frac{\al^2}{JS^2}},
\label{betaPresult}
\end{align}
where $F=2.315$ considering high temperature ($\beta\gap\ll1$).
Note that there is a minus sign in the magnon spin-transfer term (proportional to $P_T$) in Eq. (\ref{DWeq}) indicating that the magnon spin-transfer pushes the wall to the hotter end of the system as has been noted previously \cite{Hinzke11,Yan12}.
Equation (\ref{betaPresult}) indicates a clear distinction between the magnon spin-transfer effect from the electron spin-transfer effect in metals, namely, the spin-transfer efficiency parameter, $P_T$, grows at small damping ($\alpha$), while the parameter is independent on $\alpha$ in the case of electron spin-transfer effect.

The equations of motion, Eq. (\ref{DWeq}), has the same form as the current-driven case in metals.
As was shown in Ref. \cite{TTKSNF06}, the behavior of the wall are distinct at small and large $\uT$ compared to the crossover velocity (the Walker's breakdown velocity), defined as
\begin{align}
  \ua\equiv -\frac{\vc}{P_T+\frac{\beta_T}{\alpha}},
\end{align}
which depends on the temperature and $\alpha$.
At small driving velocity, $|\uT|\leq |\ua|$, the tilting angle of the wall, $\phi$, reaches the terminal angle determined by
\begin{align} 
  \sin 2\phi &= \lt(P_T+\frac{\beta_T}{\alpha} \rt) \frac{\uT}{\vc},
\end{align}
and the terminal velocity of the wall is a constant, 
\begin{align} 
  \vw &= \frac{\beta_T}{\alpha}\uT. \label{Vwbeta}
\end{align}
The wall thus moves from the high- to low-temperature region in this case, due to a pressure on the wall.
Above the crossover velocity, $|\uT|> |\ua|$, the angle $\phi$ becomes time-dependent and the average wall velocity is
\begin{align} 
  \vw &= \frac{\beta_T}{\alpha}\uT-\frac{\vc}{1+\alpha^2}
   \sqrt{\lt(P_T+\frac{\beta_T}{\alpha}\rt)^2\lt(\frac{\uT}{\vc}\rt)^2-1}.\label{Vdwres}
\end{align}
For $\uT\gg|\ua|$, $\vw = \frac{1}{1+\alpha^2}(-P_T+\alpha\beta_T)\uT$.
If $P_T-\alpha \beta_T>0$, the direction of domain wall motion therefore changes around $\uT\sim O(|\ua|)$; 
at small $\nabla T$, the motion is towards the colder end while it is opposite in  large  $\nabla T$ region, dominated by magnon spin-transfer effect.
At $\uT\gg|\ua|$, the wall velocity is 
$\vw/\nabla T=-\frac{\kB\al^2}{\hbar}(P_T-\alpha\beta_T)$. %\simeq 7\times 10^{-7}$ m$^2$/(sK).

\begin{figure}[tb]\centering
\includegraphics[width=0.3\hsize]{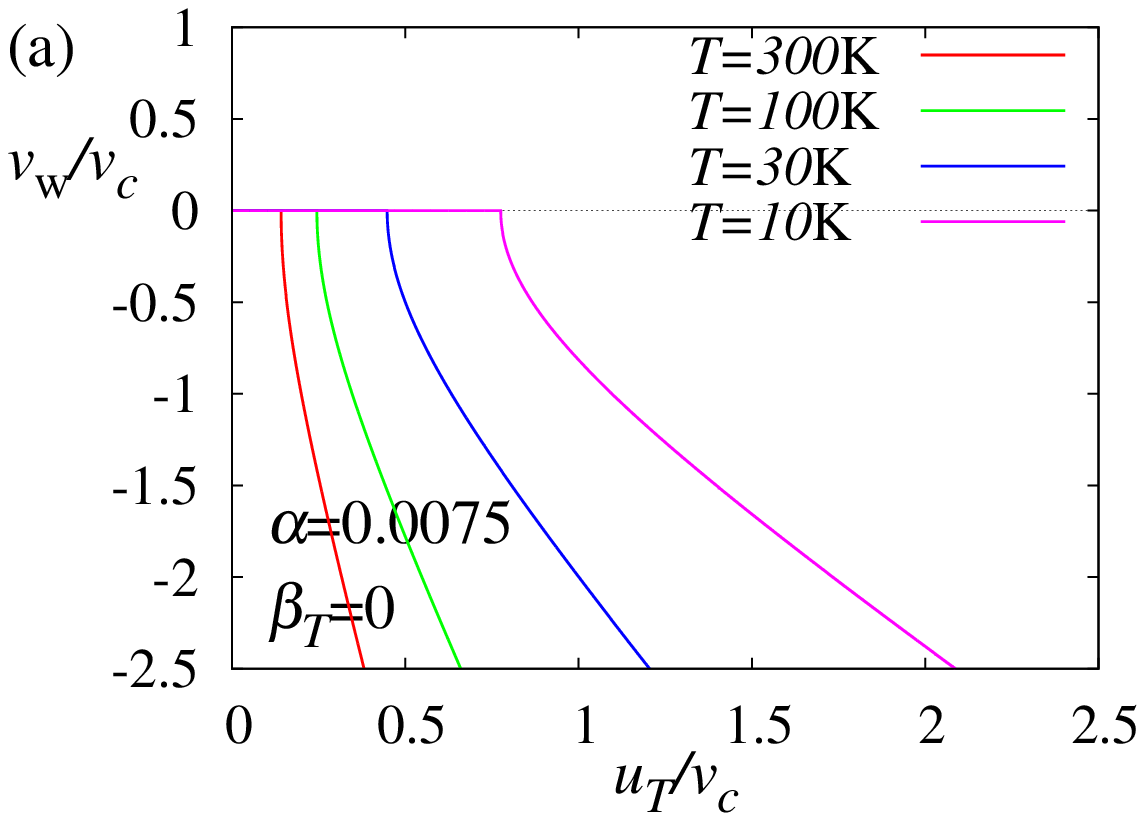}
\includegraphics[width=0.3\hsize]{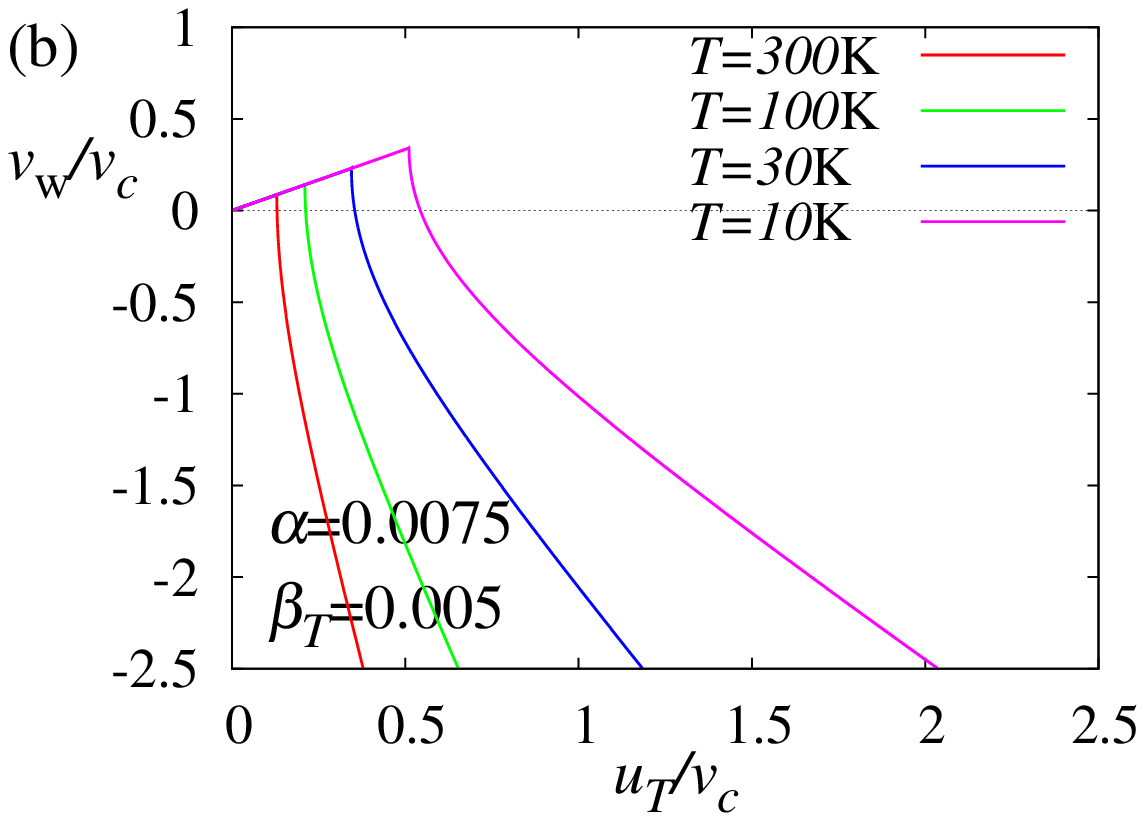}
\includegraphics[width=0.3\hsize]{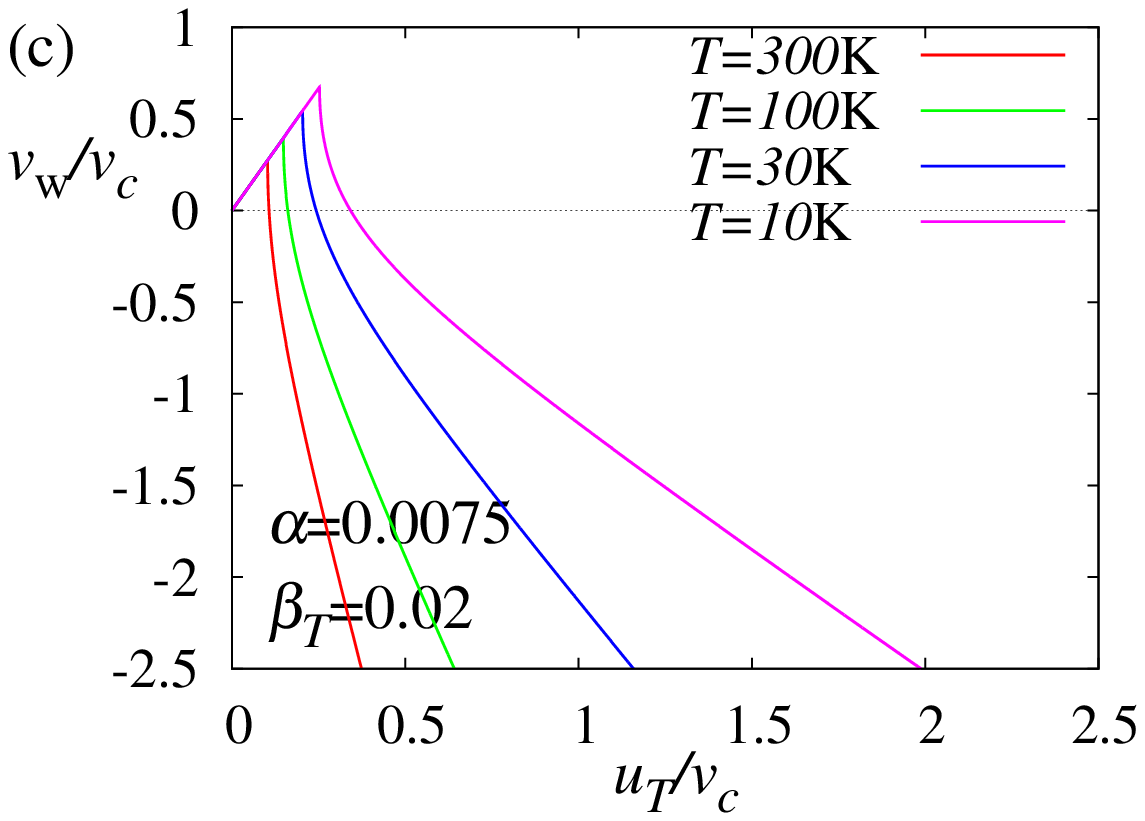}
\includegraphics[width=0.3\hsize]{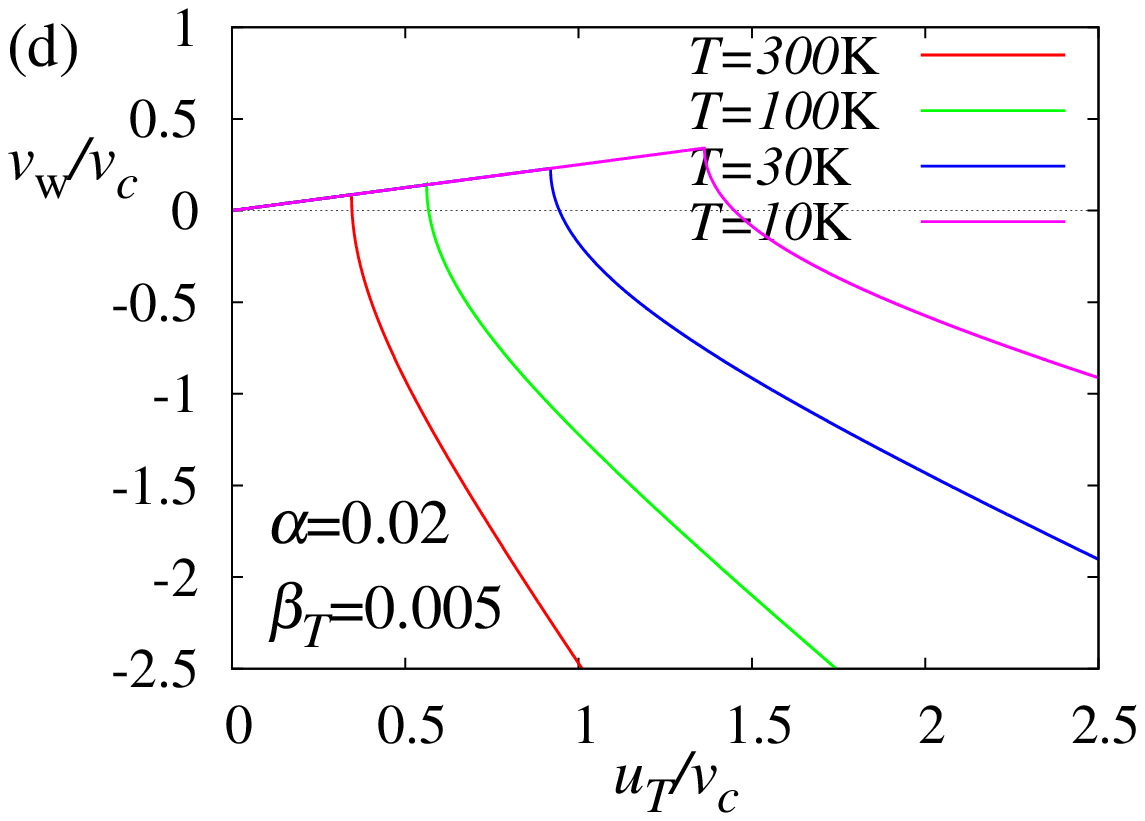}
\caption{ (a-c): The domain wall velocity in the unit of $\vc$ as a function of $\uT/\vc$ for different temperatures and for $\beta_T=0, 0.005$ and $0.02$ at  $\alpha=0.0075$.
Negative $\vw$ is towards the direction to the hotter end. 
Parameters used are $\al=12$\AA, $S=14$, $\frac{JS^2}{\al^2\kb}=4.2\times 10^2$ K. 
The case of $\alpha=0.02$ with $\beta_T=0.005$ is plotted as (d).
Effects of extrinsic pinning is not taken into account.
It is seen that the magnon spin-transfer effect is enhanced at high temperature and when $\alpha$ is small.
\label{FIGVdw}}
\end{figure}

The domain wall velocity as a function of $\uT/\vc$ is plotted in Fig. \ref{FIGVdw} for different temperatures and $\beta_T$.
When there is no force ($\beta_T=0$), there is a threshold value of $\frac{\vc}{P_T}$ for $u_T$ for thermal motion to set in, and the motion towards the hotter regime occurs above the threshold.
The wall speed increases as function of the temperature, since the effective spin polarization, $P_T$, increases at high temperatures rather significantly ($\propto \sqrt{T}$).  
If $\beta_T$ is positive, wall motion towards the colder regime occurs at small $u_T(<|\ua|)$, and the magnon spin-transfer regime appears at $|u_T|\gtrsim|\ua|$. 
For large $\alpha$, the wall speed is smaller (Fig. \ref{FIGVdw}(d)).

%%%%%%%%%%%%%%%%%
%%%%%%%%%%%%%%%%%

%%%%%%%%%%%%%%%%%
\begin{figure}[tb]\centering
\includegraphics[width=0.3\hsize]{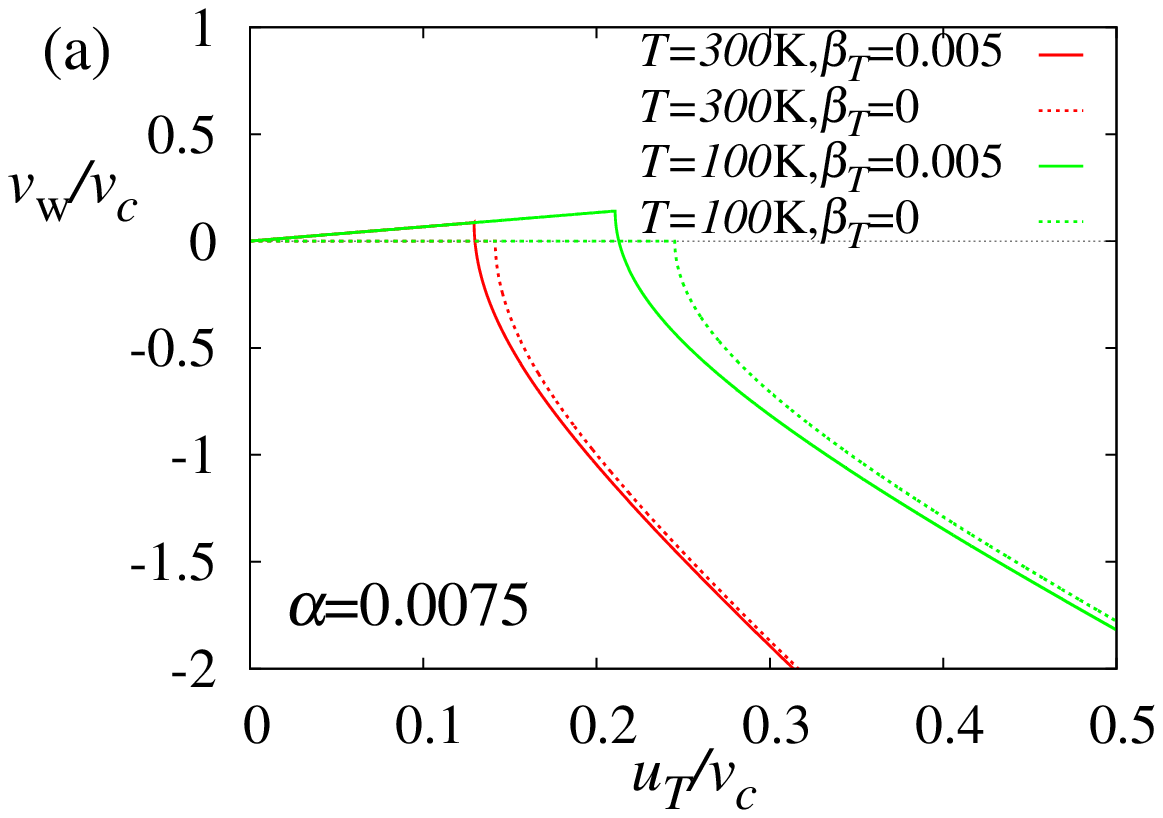}
\includegraphics[width=0.3\hsize]{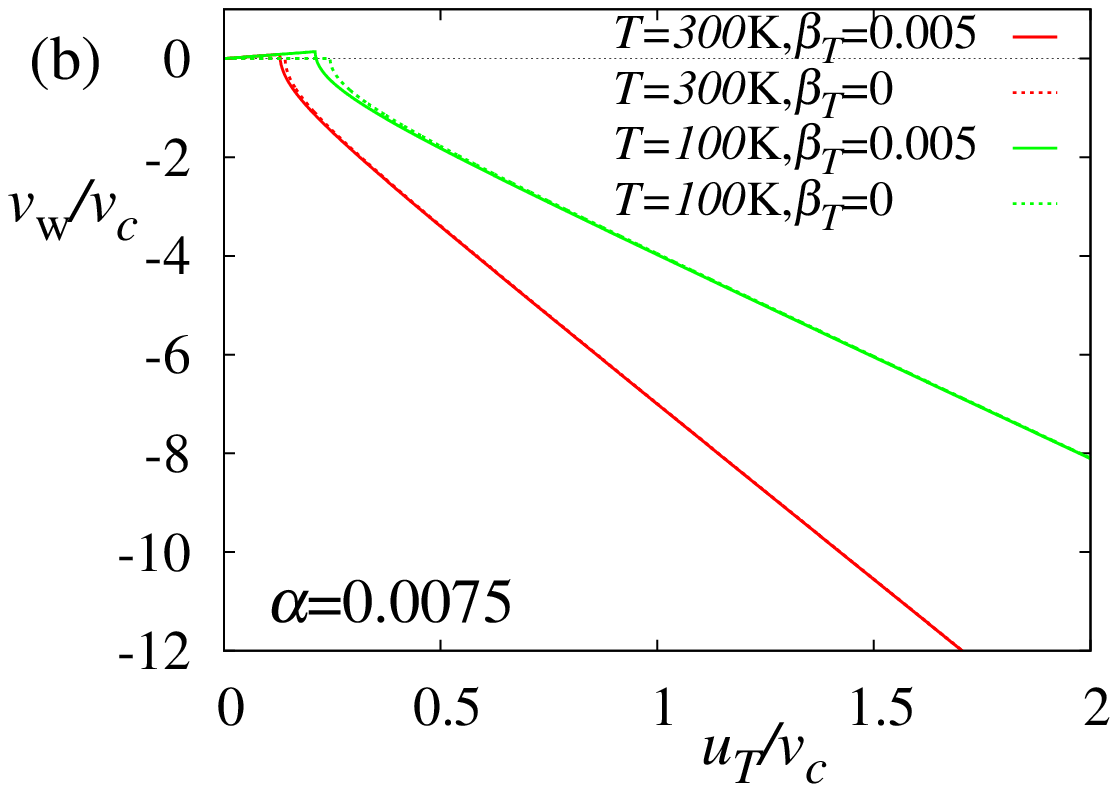}
\caption{ The domain wall velocity for $\beta_T=0$ and $\beta=0.005$ at $T=100$K and $300$K at  $\alpha=0.0075$ for (a) small and (b) large $\uT/\vc$. We see that dependence on value of $\beta_T$ is weak.
\label{FIGVdwLow}}
\end{figure}
%%%%%%%%%%%%%%%%%

In YIG, $\alpha=0.0075$, $\al=12$\AA, and the magnetization is $M=1.4\times 10^{5}$ A/m.
If we convert the magnetization to the average magnitude of spin by $M=\frac{2\mub}{a^3}S$, we obtain the effective spin of $S=14$.
The exchange stiffness is $A=4.7\times 10^{-12}$ J/m, which corresponds to 
$\frac{JS^2}{\al^2}=A\al=5.8\times10^{-21}$ J ($\frac{JS^2}{\al^2\kb}=4.2\times 10^2$ K).
In the three-dimensional case, we therefore have $P_T=0.40\times\sqrt{T({\rm K})}$. 
(At $T=300$ K, $P_T=6.7$ and the crossover velocity is $\ua=-\vc\times 0.15$.)
The absolute value of the wall speed depends on the hard-axis anisotropy energy, $\Kperp$ ($\vc\propto \Kperp$).
The easy-axis anisotropy energy of YIG is $KS^2\al^3=1.2\times 10^{-24}$ J, and 
the hard-axis energy is much weaker than the easy-axis one. 
Let us choose $\Kperp/K= 10^{-3}$, i.e., $\vc= 2.4\times 10^{-2}$ m/s.
We consider the case  $\beta_T=0$ first. 
We have $|\ua|=0.36\times 10^{-2}$ m/s at $T=300$ K.
In this case, 
a temperature gradient of $\nabla T=20$ K/mm in Ref. \cite{Jiang13} corresponding to  
$\uT =3.8\times 10^{-3}$ m/s ($\uT/\vc=0.16$) is close to the threshold ($\uT\simeq|\ua|$), and the wall speed is read from Fig. \ref{FIGVdw}(a) and Fig. \ref{FIGVdwLow}(a) as 
$\vw=-0.37\vc=-0.89\times 10^{-2}$ m/s.
In the case of $\beta_T=0.005$, $\uT/\vc=0.16$ leads to 
$\vw=-0.50\vc=-1.2\times 10^{-2}$ m/s (Fig. \ref{FIGVdw}(b) and Fig. \ref{FIGVdwLow}(a) ), and we see that the order of magnitude of the speed does not depend strongly on $\beta_T$ (Fig. \ref{FIGVdwLow}(a)).
An interesting fact is that the wall velocity changes sign as we lower the temperature if $\beta_T$ is finite. 
In fact, when we set $\uT/\vc=0.16$ for $\beta_T=0.005$, the wall velocity is negative at $T=300$ K, while it is positive at $T=100$ K as  seen in Fig. \ref{FIGVdwLow}(a). 
The sign change occurs around $T=190$ K, where $\ua(T)\sim \uT$. 

If hard-axis anisotropy is weaker, $\Kperp/K= 10^{-4}$,
$\vw=-11\vc=-2.6\times 10^{-2}$ m/s at $T=300$ K and $\vw=-6\vc=-1.4\times 10^{-2}$ m/s at $T=100$ K
for $\uT =3.8\times 10^{-3}$ m/s and for both $\beta_T=0$ and $0.005$ 
(Fig. \ref{FIGVdwLow}(b)).

In the experiment on YIG at the room temperature in Ref. \cite{Jiang13}, the wall moved to the hotter regime at speed of $1.8\times 10^{-4}$ m/s at $\nabla T=22$ K/mm.
This speed is smaller than our theoretical estimate by two orders of magnitude, and we suspect that the small value would be due to the extrinsic pinning. 
In fact, the experiment was carried out under rather strong field of $B=6\times 10^{-3}$ T (60 Oe), which adds a large additional force term $\frac{g\mub B}{\hbar}=1\times 10^9$ s$^{-1}$ to the right-hand side of Eq. (\ref{DWeq1}). This force is expected to have been necessary to compensate extrinsic pinning effects, and this fact and the avalanche behavior of walls observed suggest a strong influence of extrinsic pinning in their sample. 

In the experiment carried out in metals in Ref. \cite{Jen_b86}, the applied $\nabla T$ was order of 100 K/mm. The motion was observed in the presence of ac magnetic field to remove the extrinsic pinning effects and wall velocity of 2 mm/s towards the colder end was obtained. 
The direction would be understood by the spin-transfer effect of thermally-induced  conduction electron flow.

\section{Conclusion}

To summarize, we applied a thermal vector potential theory developed in Ref. \cite{Tatara15} to describe magnetization dynamics in ferromagnetic insulator driven by a temperature gradient.
We have evaluated the magnon current induced by the temperature gradient within the linear response theory and it was found to be proportional to the inverse of the Gilbert damping parameter.
The effect of magnon current is thus  dominant in weak damping systems as was argued in Ref. \cite{Kong13}.
The magnon current was shown to   exerts a spin-transfer torque on a domain wall, and this effect tends to drive the structure towards the hotter side.
The case of a vortex (or a skyrmion) was also discussed.

\acknowledgements
The author  thanks  H. Kohno  and S.-K. Kim
for valuable discussions.
This work was supported by a Grant-in-Aid for Scientific Research (C) (Grant No. 25400344), (A) (Grant No. 24244053) from Japan Society for the Promotion of Science and  
Grant-in-Aid for Scientific Research on Innovative Areas (Grant No.26103006) from The Ministry of Education, Culture, Sports, Science and Technology (MEXT), Japan.

%%%%%%%%%%%%%%%%%%%%%%%%%%%%%%%%%%%%%%%%%%%%%%%%%%%%%%%%
%%%%%%%%%%%%%%%%%%%%%%%%%%%%%%%%%%%%%%%%%%%%%%%%%%%%%%%%
%%%%%%%%%%%%%%%%%%%%%%%%%%%%%%%%%%%%%%%%%%%%%%%%%%%%%%%%%%%%%%%%%%%%%%%%%%%%%%%%%%%%%%%%%%%
\appendix
%%%%%%%%%%%%%%%%%%%%%%%%%%%%%%%%%%%%%%%%%%%%%%%%%%%%%%%%
%%%%%%%%%%%%%%%%%%%%%%%%%%%%%%%%%%%%%%%%%%%%%%%%%%%%%%%%

\section{Energy current density of localized spin \label{SECJEspin} }

We derive here the expression for the energy current density of the localized spin by evaluating the time-derivative of the energy density quantum mechanically.
The Hamiltonian we consider is the one with the exchange interaction and easy and hard axis anisotropy energies, Eq. (\ref{HamiltonianDW}).
We carry out the calculation on a discretized lattice, since the estimation of commutators goes straightforwardly, and then take the  continuum limit.
The energy densities at site $i$ are 
\begin{align}
{\cal E}^J_i &\equiv -\frac{J}{2a^5}\sum_{\alpha\beta}\sum_{\sigma=\pm} S^\alpha_i S^\alpha_{i +\sigma \beta} \nnr 
{\cal E}^K_i & \equiv - \frac{K}{2a^3} (S^z_i)^2 , \;\;\; 
{\cal E}^{K_\perp}_i  \equiv \frac{\Kperp}{2a^3} (S^y_i)^2
   ,   \label{spinenergydensities}
\end{align}
where $\alpha$, $\beta$ runs $x,y$ and $z$, and $\sigma$ denotes the positive and negative directions.
Corresponding terms of discretized Hamiltonian are 
\begin{align}
 H_J & \equiv a^3\sum_i {\cal E}^J_i , \;\;\; 
 H_K \equiv a^3\sum_i {\cal E}^K_i,  \;\;\; 
 H_{K_\perp} \equiv a^3\sum_i {\cal E}^{K_\perp}_i.
\end{align}

We first study the exchange interaction contribution by evaluating $[{\cal E}_i^J,H_J]$;
\begin{align}
   [{\cal E}_i^J,H_J] &= \frac{J^2}{4a^7}\sum_j \sum_{\alpha\alpha'\beta\beta'}\sum_{\sigma\sigma'}
  [S^\alpha_i S^\alpha_{i +\sigma\beta},S^{\alpha'}_j S^{\alpha'}_{j+\sigma'\beta'}].
\end{align}
By use of $[S_i^\alpha,S_j^\beta]=i\delta_{ij}\epsilon_{\alpha\beta\gamma} S_i^\gamma$ and 
\begin{align}
  [AB,CD]=A[B,C]D+[A,C]BD+CA[B,D]+C[A,D]B,
\end{align}
we see that 
\begin{align}
 \sum_j & [S^\alpha_i S^\alpha_{i +\sigma \beta},S^{\alpha'}_j S^{\alpha'}_{j+\sigma'\beta'}]
 \nnr
 & = 
i\sum_{\delta} 
\epsilon_{\alpha\alpha'\delta} 
         \lt(  S^\alpha_i  S^\delta_{i +\sigma \beta} S^{\alpha'}_{i +\sigma \beta+\sigma'\beta'}
      +  S^\delta_i S^\alpha_{i +\sigma \beta} S^{\alpha'}_{i+\sigma'\beta'} \rt. \nnr
  & \lt. 
      + S^{\alpha'}_{ i +\sigma \beta-\sigma'\beta'} S^\alpha_i S^\delta_{i +\sigma \beta}
      + S^{\alpha'}_{i-\sigma'\beta'} S^\delta_i S^\alpha_{i +\sigma \beta} \rt) \nnr
       & =
i\sum_{\delta} 
\epsilon_{\alpha\alpha'\delta} 
        \biggl(  
        -2 S^\alpha_i  S^{\alpha'}_{i +\sigma \beta} 
           (S^{\delta}_{i +\sigma \beta+\sigma'\beta'}-S^{\delta}_{i +\sigma'\beta'})   \nnr
  & 
      +  \sum_{\epsilon} \epsilon_{\alpha\alpha'\epsilon} 
        (\delta_{\sigma\beta,\sigma'\beta'} S^\epsilon_i S^\delta_{i +\sigma \beta} 
           -\delta_{\sigma\beta,-\sigma'\beta'} S^\delta_i S^\epsilon_{i +\sigma \beta} ) \biggr) ,
\end{align}
where we used the fact that spins on different sites commute each other.
The contribution from the last two terms vanish after summation over $\sigma$ and $\sigma'$, and we obtain 
\begin{align}
 [{\cal E}_i^J,H_J] &= -i\frac{J^2}{2a^7} \sum_{\alpha\alpha'\beta\beta'\delta}\sum_{\sigma,\sigma'=\pm}
\epsilon_{\alpha\alpha'\delta} 
 S^\alpha_i  S^{\alpha'}_{i +\sigma \beta}  \nnr
 & \times
           (S^{\delta}_{i +\sigma \beta+\sigma'\beta'}-S^{\delta}_{i +\sigma'\beta'}).
 \label{JJcomres}
\end{align}
The commutators including the exchange interaction and easy-axis anisotropy energy are 
\begin{align}
 [&{\cal E}_i^J,H_K] + [{\cal E}_i^K,H_J]
   = -i\frac{JK}{4a^5} \sum_{\alpha\alpha'\beta}\sum_{\sigma=\pm}
\epsilon_{\alpha\alpha' z} 
 \biggl[ 
  S_i^\alpha  S^{\alpha'}_{i +\sigma \beta}S_{i +\sigma \beta}^z \nnr
 &
 + S_{i +\sigma \beta}^z S_i^\alpha  S^{\alpha'}_{i +\sigma \beta}
 -S_{i+\sigma \beta}^\alpha  S^{\alpha'}_{i}S_{i}^z
 -S_{i}^z S_{i+\sigma \beta}^\alpha  S^{\alpha'}_{i} \biggr].
 \label{JKcomres2}
\end{align}
The hard-axis anisotropy contribution has the same form but with the index $z$ replaced by $y$.

Let us take the continuum limit.
Contribution from the exchange interaction is ($\mu=x,y,z$) 
\begin{align}
 [{\cal E}_i^J,H_J] 
 &= 
 i\frac{J^2}{2a^3} \sum_{\alpha\alpha'\beta\delta}\sum_{\sigma=\pm}
\epsilon_{\alpha\alpha'\delta}  
  \lt(\sigma \nabla_\beta S^{\alpha} \rt) S^{\alpha'} 
\sigma\nabla_\beta  (\nabla^2 S^{\delta})
\nnr
&=
 -i\hbar  \nabla \cdot \jv_{{\cal E}}^{JJ},
  \label{JJcomcont}
\end{align}
where 
\begin{align}
  j_{{\cal E},\mu}^{JJ}
  \equiv  \frac{J^2}{\hbar a^3} \biggl[\nabla_\mu\Sv\cdot [(\nabla^2 \Sv)\times \Sv ] \biggr],
\end{align}
is the energy current from the exchange interaction.
Similarly
\begin{align}
 [{\cal E}_i^J,H_K] + [{\cal E}_i^K,H_J]
  &= -i\hbar  \nabla \cdot \jv_{{\cal E}}^{JK}
  \nonumber\\
 [{\cal E}_i^J,H_{K_\perp}] + [{\cal E}_i^{K_\perp} ,H_J]
  &= -i\hbar  \nabla \cdot \jv_{{\cal E}}^{JK_\perp},
\end{align}
where 
\begin{align}
  \jv_{{\cal E},\mu}^{JK}
  & \equiv  \frac{JK}{4\hbar a^3} 
  \biggl[(\Sv\times\nabla_{\mu}\Sv-\nabla_{\mu}\Sv\times\Sv)_z S_z \nnr 
  & +S_z(\Sv\times\nabla_{\mu}\Sv-\nabla_{\mu}\Sv\times\Sv)_z
  \biggr]\nnr
  \jv_{{\cal E},\mu}^{JK_\perp}
  & \equiv - \frac{JK_\perp}{4\hbar a^3} 
  \biggl[(\Sv\times\nabla_{\mu}\Sv-\nabla_{\mu}\Sv\times\Sv)_y S_y  \nnr
  & +S_y(\Sv\times\nabla_{\mu}\Sv-\nabla_{\mu}\Sv\times\Sv)_y
  \biggr].
\end{align}
In the classical case, where ordering does not matter, we obtain 
\begin{align}
  \jv_{{\cal E},\mu}^{JK}
  & =  \frac{JK}{\hbar a^3} 
  (\Sv\times\nabla_{\mu}\Sv)_z S_z
  = -\gyro\frac{J}{a^3}  \nabla_{\mu}\Sv\cdot(\Bv_{K}\times\Sv) \nnr
  \jv_{{\cal E},\mu}^{JK_\perp}
  & =  -\gyro\frac{J}{a^3}  \nabla_{\mu}\Sv\cdot(\Bv_{K_\perp}\times\Sv),
\end{align}
where $\gyro$ is gyromagnetic ratio and 
\begin{align}
  \hbar\gyro\Bv_{K} & \equiv - KS_z \hat{\zv} \nonumber\\
  \hbar\gyro\Bv_{K_\perp} & \equiv  K_\perp S_y \hat{\bm y} ,
\end{align}
are the effective magnetic fields for the easy and hard axis anisotropy energies, respectively
( $\hat{\bm y}$ and $\hat{\zv}$ are the unit vectors along $y$ and $z$ axis, respectively).
There is no contribution to the energy current from the commutators of anisotropy energies, since local quantities commute each other.
The total energy current for the Hamiltonian (\ref{HamiltonianDW}) is therefore 
\begin{align}
  \jv_{{\cal E},\mu} 
  &\equiv
  \jv_{{\cal E},\mu}^{JJ} + \jv_{{\cal E},\mu}^{JK} + \jv_{{\cal E},\mu}^{JK_\perp} \nonumber\\
  & = 
   - \gyro\frac{J}{a^3}  \nabla_{\mu}\Sv\cdot(\Bv_{H}\times\Sv) ,
\end{align}
where 
\begin{align}
 \hbar\gyro\Bv_{H} &\equiv \frac{\delta H}{\delta \Sv} 
   = -J \nabla^2\Sv - KS_z \hat{\zv} + K_\perp S_y \hat{\bm y},
\end{align}
is the total effective magnetic field for the Hamiltonian $H$.
The equation of motion for $\Sv$ is
\begin{align}
  \dot{\Sv}=\gyro \Bv_{H}\times\Sv,
\end{align}
and thus we finally obtain the energy current density as
\begin{align}
  \jv_{{\cal E},\mu} 
  & = 
   - \frac{J}{a^3}  \nabla_{\mu}\Sv\cdot\dot{\Sv} .\label{JEspinresult}
\end{align}
This form in fact the one obtained from a general definition in terms of the Lagrangian $L$, 
$T_{0i}\equiv (\partial_t \Sv) \frac{\delta L}{\delta \partial_i \Sv}$. 
It also agrees with the one proposed in Ref. \cite{Kovalev12} based on a symmetry argument, although the spin relaxation was believed there to be essential for emergence of a term of Eq. (\ref{JEspinresult}).

%%%%%%%%%%%%%%%%%%%%%%%%%%%%%%%%%%%%%%%%%%%%%%%%%%%%%%%%
%%%%%%%%%%%%%%%%%%%%%%%%%%%%%%%%%%%%%%%%%%%%%%%%%%%%%%%%

%%%%%%%%%%%%%%%%%%%%%%%%
%\bibliography{/home/tatara/References/14,/home/tatara/References/gt}

\end{document}